\documentclass[aps,twocolumn,showpacs,floatfix]{revtex4}

\usepackage{dcolumn}
\usepackage{graphicx}
\usepackage{amsfonts}
\usepackage{amsmath,bm,amssymb}

\begin{document}

\title{Spin moment formation and reduced orbital polarization in
  LaNiO$_3$/LaAlO$_3$ superlattice: LDA+U study}

\author{Myung Joon Han} \affiliation{ Department of Physics, Northern
  Illinois University, De Kalb, Illinois 60115, USA}
  \affiliation{Advanced Photon Source, Argonne National Laboratory,
    9700 South Cass Avenue, Argonne, Illinois 60439, USA}

  \author{ Michel van Veenendaal} \affiliation{ Department of Physics,
    Northern Illinois University, De Kalb, Illinois 60115, USA}
  \affiliation{Advanced Photon Source, Argonne National Laboratory,
    9700 South Cass Avenue, Argonne, Illinois 60439, USA}

\date{\today }

\begin{abstract}
  Density functional band calculations have been performed to study
  LaNiO$_3$/LaAlO$_3$ superlattices. Motivated by recent experiments
  reporting the magnetic and metal-insulator phase transition as a
  function of LaNiO$_3$ layer thickness, we examined the electronic
  structure, magnetic properties, and orbital occupation depending on
  the number of LaNiO$_3$ layers. Calculations show that the magnetic
  phase is stabler than the nonmagnetic for finite and positive $U$
  values. The orbital polarization is significantly reduced by $U$
  even in the magnetic regions. The implications of the results are
  discussed in comparison to recent experimental and theoretical
  studies within the limitations of the LDA$+U$ method.
\end{abstract}


\maketitle

\section{Introduction}

Understanding transition metal oxides is of perpetual interest and
importance in condensed matter physics and material science
\cite{MIT-RMP}.  Recent advances in layer-by-layer growth techniques
of heterostructures of transition metal compounds have created
particular interest due to their great scientific and technological
potential \cite{MRS}.  Exotic material phenomena that are clearly
distinctive from the `normal' phases include interface
superconductivity~\cite{Reyren}, magnetism
\cite{mag-LAOSTO-1,mag-LAOSTO-2}, charge~\cite{Ohtomo-1,Okamoto} and
orbital reconstruction~\cite{OrbReconst}.

One of the most intriguing classes of materials may be the nickelate
superlattices
\cite{LNOLAO-NatMatt,Boris-Science,MJHan-DMFT,Hansmann,MJHan,
  Liu-PRB-RC,John-arxiv,NNO-PRL}. A series of recent theoretical
studies have created considerable interest, suggesting the
heterostructuring-induced orbital polarization and the possible
high-T$_c$ superconductivity in LaNiO$_3$ (LAO)/LaAlO$_3$ (LAO)
superlattice \cite{Hansmann,Chaloupka-Khaliullin}. In this picture,
the two degenerate Ni $e_g$ orbital states are split by a combination
of translational symmetry breaking and on-site Coulomb interaction,
leading to a cuprate-like band structure. Although this kind of
theoretical picture has been challenged by more recent dynamical
mean-field theory (DMFT) calculations based on a charge transfer model
including oxygen states explicitly \cite{MJHan-DMFT}, several
experimental papers have found other interesting phenomena in this
system.  Boris and co-workers \cite{Boris-Science} reported a metal to
insulator transition as a function of LNO layer thickness: Even though
the bulk LNO is a paramagnetic (PM) metal, the heterostructure
(LNO)$_m$/(LAO)$_n$ becomes insulating and magnetic if $m$ is small,
$m\leq 2$, while it remains PM and metallic when $m\geq 4$
\cite{Boris-Science}. The insulating behavior was also observed by
Freeland {\it et al.}  \cite{John-arxiv}. However the detailed
magnetic and electronic structure changes, as well as the other
important physical quantities such as orbital polarization have not
yet been clearly understood as a function of layer thickness, $m$.

In this study, we performed a detailed first-principles analysis for
the electronic structure, magnetism, and orbital polarization of
LNO/LAO superlattice using the LDA$+U$ method
\cite{LDA+U-ori,LDA+U-95,LDA+U-rev}. Since previous calculations were
performed at the LDA \cite{MJHan} or DMFT level
\cite{Hansmann,MJHan-DMFT} and assumed the bulk-like PM phase, LDA$+U$
calculation can provide meaningful information especially regarding
the magnetism in this system. Total energy calculations showed that
the LDA+$U$ stabilizes ferromagnetic (FM) spin order along in-plane
and out-of-plane direction. This result may indicate the existence of
another ground state configuration in between $m\sim 2$ and $m\sim 4$
superlattices. The orbital polarization is significantly reduced by
$U$ which is in contrast to the simplified Hubbard-type model
prediction \cite{Hansmann}, but consistent with the extended
charge-transfer model DMFT calculation \cite{MJHan-DMFT}. The
calculated Ni-$d$ valence based on LDA$+U$ supports the recently
suggested picture for the metal/insulator phase diagram based on the
$d$-valency \cite{Wang}, and demonstrates the importance of the double
counting issue. These results are discussed in comparison to the
recent theoretical and experimental studies.

\begin{figure}[t]
  \centering \includegraphics[width=6cm]{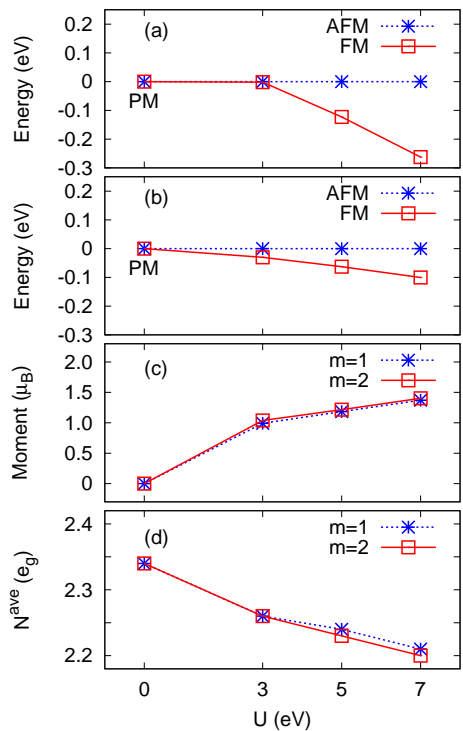}
  \caption{(Color online) (a) Calculated total energy of
    (LNO)$_1$/(LAO)$_1$ with the in-plane spin ordering of FM and AFM
    as a function of $U$.  (b) Calculated total energy of
    (LNO)$_2$/(LAO)$_1$ with the out-of-plane spin ordering of FM and
    AFM as a function of $U$ where the in-plane spins are set to be FM.
    (c) Calculated Ni magnetic moment for (LNO)$_1$/(LAO)$_1$ and
    (LNO)$_2$/(LAO)$_1$ as a function of $U$ (FM order considered). (d)
    Average number of Ni-$e_g$ electrons in (LNO)$_1$/(LAO)$_1$ and
    (LNO)$_2$/(LAO)$_1$ as a function of $U$.
    \label{EvsU}}
\end{figure}

\section{Computational Details}
For the band-structure calculations, we employed Troullier-Martins
type norm-conserving pseudopotential \cite{troullier} with a partial
core correction and linear combination of the localized pseudo-atomic
orbitals (LCPAO) \cite{Ozaki} as a basis set. In this pseudo-potential
generation, the semi-core 3$p$ electrons for transition metal atoms
were included as valence electrons in order to take into account the
contribution of the semi-core states to the electronic structure. We
adopted the local density approximation (LDA) for exchange-correlation
energy functional as parametrized by Perdew and Zunger \cite{CA}, and
used energy cutoff of 400 Ry and k-grid of $12\times 12\times 6$ per
unit superlattice volume.  The LDA$+U$ functional is adapted from the
formalism of Ref.~\onlinecite{Dudarev} and Ref.~\onlinecite{Han-LDA+U}
\cite{comment0}. The geometry relaxation has been performed with the
force criterion of $10^{-3}$ Hartree/Bohr. During the relaxation
process, the in-plane lattice constant is fixed considering the
substrate effect in the experimental situation. The orbital
polarization $P$, defined as
\begin{equation}
  P=\frac{n_{x^2-y^2}-n_{3z^2-r^2}}{n_{x^2-y^2}+n_{3z^2-r^2}},
\label{Pdef}
\end{equation}
can be calculated by integrating the projected density-of-states (DOS)
up to Fermi level.  All the calculations were performed using the
density functional theory code OpenMX \cite{OpenMX}.

\begin{figure}[t]
  \centering \includegraphics[width=8cm]{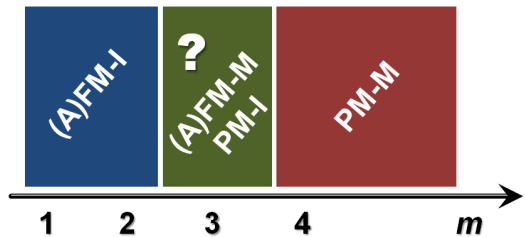}
  \caption{(Color online) The suggested schematic phase diagram
    (based on the previous experiment and our calculation) of the
    possible ground state configuration of (LNO)$_m$/(LAO)$_n$ as a
    function of $m$ (see the text).
    \label{pdia}}
\end{figure}

\section{Result and Discussion}
One natural and evidently important question raised from experimental
studies on the magnetic moment formation in this heterostructure
\cite{Boris-Science} is about the ground state spin structure and its
moment size. Fig.~\ref{EvsU} summarizes our results.
Fig.~\ref{EvsU}(a) shows the calculated total energy of
(LNO)$_1$/(LAO)$_1$ superlattice as a function of $U$. While at $U$=0
eV both FM and antiferromagnetic (AFM) spins eventually converge to a
PM solution, the magnetic ground states are stabilized at the finite
$U$. As clearly seen in Fig.~\ref{EvsU}(a), FM spin ordering is
energetically favored within the LNO plane. The energy difference
between FM and AFM state is 3, 246, and 525 meV per
(LNO)$_1$/(LAO)$_1$ for $U$=3, 5, and 7 eV, respectively. FM spin
order is also favored along the out-of-plane direction as shown in
Fig.~\ref{EvsU}(b) in which we present total energies of
(LNO)$_2$/(LAO)$_1$; the out-of-plane spin orderings are set to be FM
or AFM while the in-plane order is FM. The calculated FM-AFM energy
difference is 60, 126, and 201 meV per (LNO)$_2$/(LAO)$_1$ for $U$=3,
5, 7 eV, respectively. The energy differences between FM and AFM spin
structure becomes larger as $U$ increases in both cases of in-plane
and out-of-plane ordering. That is, the on-site correlations stabilize
FM spin ordering, and as $U$ decreases, the two magnetic solutions
becomes more close in their energies, and eventually converges to the
PM phase at $U$=0. The magnetic moment is also dependent on
$U$. Fig.~\ref{EvsU}(c) shows the calculated magnetic moment for FM
case as a function of $U$.  The moment increases as $U$ increases as
in the other typical correlated transition metal oxide materials
\cite{LDA+U-ori,LDA+U-rev,Han-LDA+U}. It is noted that, as soon as $U$
is turned on, the Ni moment becomes $\sim$ 1 $\mu_B$ ($U$=3 eV), and
further increases to be 1.2 and 1.4 $\mu_B$ at $U$=5 and 7 eV,
respectively.

\begin{figure}[t]
  \centering \includegraphics[width=6cm]{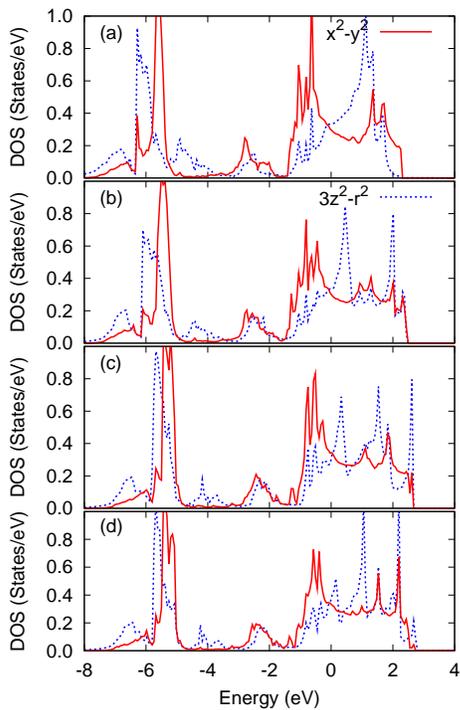}
  \caption{(Color online) Ni-$e_g$ DOS of (LNO)$_m$/(LAO)$_1$
    superlattice geometry with PM spin ($U$=0): (a) $m=1$, (b) $m=2$,
    (c) $m=3$, and (d) $m=4$. Solid (red) and dotted (blue) lines
    correspond to $d_{x^2-y^2}$ and $d_{3z^2-r^2}$ states,
    respectively, and Fermi level is set to be 0.
  \label{pdos_PM_U0}}
\end{figure}

Our calculation results may seem to suggest that the magnetic moment
formed in the thin-LNO superlattice \cite{Boris-Science} is ordered
ferromagnetically. Since the spin polarized oxide heterostructure
could be useful for the device applications, there has been active
research for finding the structure that produces FM spin order
\cite{Luders,Burton-Tsymbal,Nanda-Satpathy}. Therefore our result of a
FM ground state in LNO/LAO may have a positive implication for such an
application.  However it should be noted that the muon spin rotation
($\mu$SR) experiment by Boris {\it et al.} \cite{Boris-Science} is not
well interpreted in the long range FM ordering picture even though the
$\mu$SR is basically a local probe and that the origin of
metal-insulator phase transition in the nickelate series are not
clearly understood yet. Especially regarding the charge disproportion
or ordering in nickelates, the conventional LDA$+U$ has a clear
limitation to describe such phenomena \cite{Medarde,Mazin}. Moreover
Fig.~\ref{EvsU}(a) and (b) indicate the possibility of AFM ground
state in the negative $U$ region which is more or less related to the
reported charge disorders in the nickelate systems
\cite{Medarde,Mazin}. Therefore one needs to be careful in the
interpretation of our LDA$+U$ results on the FM spin ground state as
an indication of long range ordered ground state as one may see in the
actinide systems for example \cite{RMP-actinide,MJHan-PuAm}.

\begin{figure}[t]
  \centering \includegraphics[width=6cm,angle=0]{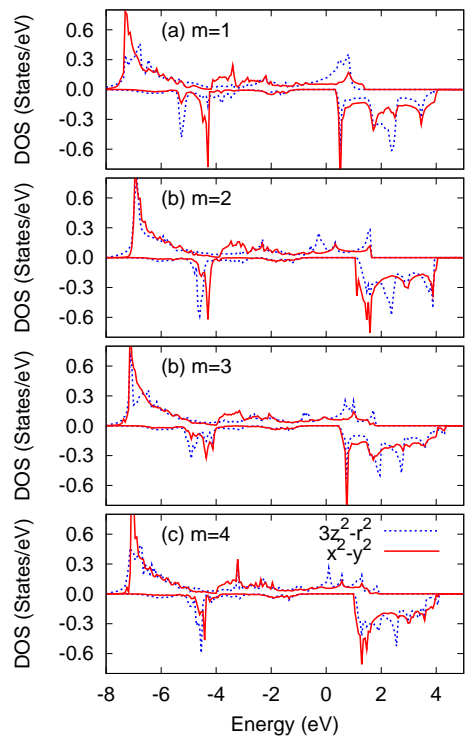}
  \caption{(Color online) Ni-$e_g$ DOS of (LNO)$_m$/(LAO)$_1$
    superlattice geometry with FM spin ($U$=5 eV): (a) $m=1$, (b)
    $m=2$, (c) $m=3$, and (d) $m=4$. Up/down panels represent
    the up/down spin states, respectively. Solid (red) and
    dotted (blue) lines correspond to $d_{x^2-y^2}$ and $d_{3z^2-r^2}$ 
    states, respectively, and the Fermi level is set to be 0.
    \label{pdos_FM_U5}}
\end{figure}

Our results have another interesting implication regarding the phase
diagram. Since the thin-LNO superlattice, (LNO)$_{m\leq 2}$/(LAO)$_n$,
can have either FM insulating (FM-I) ground state as predicted by
LDA$+U$ calculations, or, AFM insulating (AFM-I) one which is more
consistent with the $\mu$SR experiment, the system would have FM
metallic (FM-M) or AFM metallic (AFM-M) region in between the thin-LNO
limit ($m \leq 2$), and the bulk-like thick-LNO limit ($m \geq 4$; PM
and metallic, PM-M) \cite{Boris-Science}. One may also expect the PM
insulating (PM-I) phase stabilized in the same region of $m$; the
intermediate regime, $2 \leq m \leq 4$ (see, Fig.~\ref{pdia}). As
LDA$+U$ method is unable to describe correlated PM solutions properly
\cite{DMFT96,DMFT06}, the further pursue along this line is beyond the
scope of our study. For that purpose, one may resort the dynamical
mean field theory (DMFT) calculations with charge self-consistency
\cite{DMFT96,DMFT06}, and compare the total energy for PM, FM and AFM
configurations as we did in this study within LDA$+U$ scheme.

To understand metal-insulator phase transition as a function of LNO
thickness \cite{Boris-Science}, we examined Ni $e_g$ DOS depending on
the thickness, $m$.  The $d$-band width change as a function of $m$
may be an useful information as $U/W$ plays an important role in the
metal-insulator phase transition of rare-earth nickelate
\cite{DDSarma}.  Fig.~\ref{pdos_PM_U0} shows DOS of the PM case with
$U=0$. Even if the bulk LNO locates at the vicinity of metal-insulator
phase boundary and exhibits the correlated electron behaviors, the
electronic structure of PM LNO has been reasonably well described
within LDA (or GGA) as shown in the previous studies
\cite{DDSarma,MJHan,MJHan-slab}.  Fig.~\ref{pdos_PM_U0}(a)-(d) presents
the evolution of $e_g$ DOS as a function of $m$. It is noted that the
$d_{3z^2-r^2}$ band width notably changes, whereas the $d_{x^2-y^2}$
width remains almost same across $m$=2--4. The $m=2$ case
(Fig.~\ref{pdos_PM_U0}(b)), which was reported to be magnetic and
insulating \cite{Boris-Science}, the right edge of the $d_{3z^2-r^2}$
state is fairly similar to that of $d_{x^2-y^2}$ whereas the left-edge
is reduced. It is not certain however that such a relatively small
difference in terms of the effective band width is responsible for the
metal-insulator phase transition between $m=2$ and 4 as reported in
experiment \cite{Boris-Science}.

A complementary picture can be provided by LDA$+U$ calculations for
the magnetic phase. Fig.~\ref{pdos_FM_U5}(a)-(d) shows the evolution
of $e_g$ DOS as a function of $m$ for the FM (LNO)$_1$/(LAO)$_1$. Once
again the notable change is found in $d_{3z^2-r^2}$ band; for $m=1$,
the up-spin $d_{3z^2-r^2}$ state (upper panel) forms a fairly
localized DOS around Fermi level. This state becomes more and more
delocalized as $m$ increases as seen in
Fig.~\ref{pdos_FM_U5}(b)-(d). Once again, however, the amount of the
effective band width change depending on $m$ does not seem to be
enough to make metal-insulator phase transition across $m$=2--4. Other
origins than the simple $U/W$ change may be more relevant to the
transition \cite{Mazin,Wang}.

\begin{figure}[t]
  \centering \includegraphics[width=5cm,angle=270]{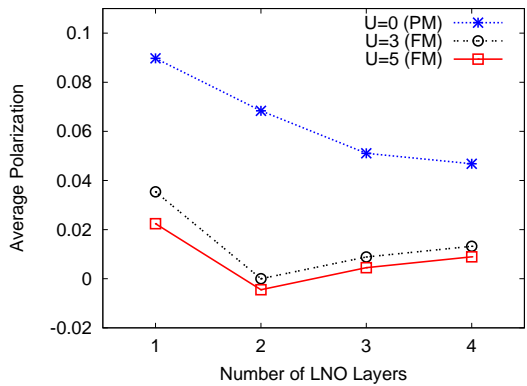}
  \caption{(Color online) The averaged orbital polarization as a 
  function of LNO layer thickness. Dotted (blue, double cross),
  double-dotted (black, circle), and solid (red, square) lines
  represent $U$=0 (PM), $U$=3 (FM), and $U$=5 (FM), respectively.
  \label{Pvsm}}
\end{figure}

An interesting point observed in Fig.~\ref{pdos_FM_U5} is that even in
LDA$+U$ calculation, the systems do not become a perfect insulator,
but have finite number of states around Fermi level
\cite{comment1}. It may partly be attributed to the strong covalency
between Ni-$d$ and O-$p$ states. As O-$p$ is much less affected by
$U$, the small amount of DOS is not removed perfectly. Another
important factor is the double-counting energy correction for which
several functional forms have been suggested
\cite{LDA+U-ori,LDA+U-95,LDA+U-rev,Han-LDA+U,Dudarev}, but there is no
well defined solution still yet \cite{Wang}: As the double-counting
energy is typically represented by $\frac{1}{2}UN_{d}(N_{d}-1)$, the
LDA$+U$ charge self-consistency adjusts $N_d$ (or, the effective
$d$-level energy and therefore the charge transfer energy,
$\epsilon_p-\epsilon_d$), depending on $U$. According to the recent
DMFT study, which tunes the double-counting term as a parameter, the
system eventually becomes insulating at large $U$
\cite{MJHan-DMFT,comment2}. Therefore the small states around the
Fermi level obtained by LDA$+U$ can be attributed to the
double-counting error that is hardly handled within the current
formalism of LDA$+U$.

An interesting recent finding on (3-dimensional) nickelates is that
the metal-insulator transition occurs at the very narrow region of
$N_d$ and the same holds for the cuprates \cite{Wang}. From the
single-site DMFT calculations with a double-counting energy as another
tuning parameter, Wang {\it et al.}  \cite{Wang} showed that there is
a well defined $N_c$, that is, the critical value of $d$-valency of
transition metal; the system is metallic if $N_d \geq N_c$ and
insulating if $N_d \leq N_c$. For nickelates, $N_c^{e_g} \approx 1.3$.
That is, for the parameters (implicitly including double-counting
correction), which result in $N_d \geq N_c$, the system remains
metallic even for the very large $U$. This conclusion suggests $N_d$
as the critical variable for understanding charge-transfer systems
\cite{Wang}. Now it might be instructive to analyze our LDA$+U$
results within this new picture of Ref.~\onlinecite{Wang}. Even if
LDA$+U$ is a lower level approximation compared to DMFT, a big merit
of LDA$+U$ is that it can be performed with the whole charge
self-consistency and take magnetism into account while the previous
DMFT calculations have dealt with the PM phase
\cite{MJHan-DMFT,Hansmann}.  Fig.~\ref{EvsU}(d) presents $N_d$ for PM
($U$=0) and FM ($U$=3, 5, 7) calculations of (LNO)$_{m=1}$/(LAO)$_1$
and (LNO)$_{m=2}$/(LAO)$_1$. While $N_d$ decreases as $U$ increases,
the difference is not significant; $\Delta N_d\approx 0.15$. Therefore
the current implementation of LDA$+U$ and the charge self-consistency
roughly follow the constant $N_d$ line. Now we note that the value of
$N_d^{e_g} \geq 2.2$ is far from the $N_c \approx 1.3$ predicted by
DMFT. Therefore it is consistent with the metallic ground state and
supports the conclusion of Ref.~\onlinecite{Wang}.

Relative orbital occupation in the two $e_g$ orbitals is a central
quantity in understanding LNO/LAO systems
\cite{LNOLAO-NatMatt,MJHan,Hansmann,MJHan-DMFT}. The orbital
polarization can be non-zero due to the translation symmetry breaking
induced by heterostructuring while the bulk polarization is 0.
According to the previous calculations, highly polarized $P$ can
possibly drives the system to be high-T$_c$ superconductor
\cite{Chaloupka-Khaliullin,Hansmann}. On the other hand, a more recent
DMFT calculation based on the realistic model hamiltonian including
oxygen orbitals predicted that the polarization is actually reduced by
the on-site correlation \cite{MJHan-DMFT}. Since both of the previous
DMFT calculations assumed PM phase and did not consider the full
charge self-consistency, LDA$+U$ calculation can give a complimentary
information for the magnetic solution in spite of its limitation in
describing correlation effect compared to DMFT. The result is
presented in Fig.~\ref{Pvsm}. It is noted that the inclusion of $U$
reduces the polarization significantly. There is a large separation
between $U$=0 and $U$=3 results, while the differences between $U$=3
and 5 results are small. Importantly the calculated polarizations for
finite $U$ are order-of-magnitude same compared with the recent DMFT
results \cite{MJHan-DMFT}.  Therefore it supports the conclusion of
the recent DMFT calculation \cite{MJHan-DMFT}. The small polarization
at the finite $U$ can also be seen in DOS presented in
Fig.~\ref{pdos_FM_U5} where no big difference can be found between the
two $e_g$ orbital occupations. We note that the reduced orbital
polarization by $U$ can be compatible with the ferromagnetic spin
order. For cuprates, for example, the fully polarized $x^2-y^2$
orbital is directly related to the in-plane AFM spin order. In the
nickelates, we have one more orbital degree of freedom available and
the spin can align ferromagnetically. And actually it is found that
the enhanced ferromagnetic trend by increasing $U$ corresponds to the
reduced orbital polarization as shown in Fig.~\ref{Pvsm}.

\section{Summary}
Using the band structure calculations based on LDA$+U$, we examined
the magnetic moment, electronic structure, $d$-valence, and orbital
polarization as a function of $U$ and $m$ (LNO thickness). The
calculated results clearly showed the formation of magnetic moment at
the finite $U$ region being consistent with a recent $\mu$SR, but the
long range ordering pattern is not so clear considering the
experiment. While $d_{3z^2-r^2}$ band width is reduced as $m$
approaches to 1, it may not be enough to be responsible for the
metal-insulator transition. The calculated orbital polarization is
significantly reduced by $U$, strongly supporting the conclusion of a
recent DMFT calculations and indicating the absence of high
temperature superconductivity in this system.

\section{Acknowledgments}
This work was supported by the U.S. Department of Energy (DOE),
DE-FG02-03ER46097, and NIU’s Institute for Nanoscience, Engineering,
and Technology. Work at Argonne National Laboratory was supported by
the U.S.  DOE, Office of Science, Office of Basic Energy Sciences,
under Contract No. DE-AC02-06CH11357.


\begin{thebibliography}{99}


\bibitem{MIT-RMP} M. Imada, A. Fujimori, and Y. Tokura,
    Rev. Mod. Phys. {\bf 70}, 1039 (1998).


\bibitem{MRS} For a reveiw, see,  J. Mannhart, D. H. A. Blank,
        H. Y.  Hwang, A. J. Millis, and  J. M. Triscone,
    Bulletin of the Materials Research Society
    {\bf 33}, 1027 (2008).


\bibitem{Reyren} N. Reyren, S. Thiel, A. D. Caviglia,
  L. F. Kourkoutis, G. Hammerl, C. Richter, C. W. Schneider, T. Kopp,
  A.-S. R{\" u}etschi, D. Jaccard, M. Gabay, D. A. Muller,
  J.-M. Triscone, J. Mannhart, 
  Science {\bf 317}, 1196 (2007).

\bibitem{mag-LAOSTO-1} L. Li, C. Richter, J. Mannhart, and R. C. Ashoori,
         Nature Phys. {\bf 7}, 762 (2011).


\bibitem{mag-LAOSTO-2} J. A. Bert, B. Kalisky, C. Bell, M. Kim, Y. Hikita, H. Y. Hwang, and K. A. Moler,
         Nature Phys. {\bf 7}, 767 (2011).


\bibitem{Ohtomo-1} A. Ohtomo, D. A. Muller, J. L. Grazul, and H. Y. Hwang,
         Nature {\bf 419}, 378 (2002).


\bibitem{Okamoto} S. Okamoto and A. J. Millis,
         Nature {\bf 428}, 630 (2004).


\bibitem{OrbReconst} J. Chakhalian, J. W. Freeland, H.-U. Habermeier,
  G. Cristiani, G. Khaliullin, M. van Veenendaal, and B. Keimer,
  Science {\bf 318}, 1114 (2007).








\bibitem{LNOLAO-NatMatt} E. Benckiser, M. W. Haverkort,
   S. Brück, E. Goering, S. Macke, A. Frañó, X. Yang, O. K. Andersen, 
   G. Cristiani, H.-U. Habermeier, A. V. Boris, I. Zegkinoglou,
   P. Wochner, H.-J. Kim, V. Hinkov and B. Keimer,
         Nature Mater.  {\bf 10}, 493 (2011).

\bibitem{Boris-Science} A. V. Boris, Y. Matiks, E. Benckiser,
  A. Frano, P. Popovich, V. Hinkov, P. Wochner, M. Castro-Colin,
  E. Detemple, V. K. Malik, C. Bernhard, T. Prokscha, A. Suter,
  Z. Salman, E. Morenzoni, G. Cristiani, H.-U. Habermeier, and
  B. Keimer,
  Science {\bf 332}, 937 (2011). 

\bibitem{Liu-PRB-RC} J. Liu, S. Okamoto, M. van Veenendaal, M. Kareev,
  B. Gray, P. Ryan, J. W. Freeland, and J. Chakhalian,
  Phys. Rev. B. {\bf 83} 161102, (2011).

\bibitem{Hansmann} P. Hansmann, X. Yang, A. Toschi, G. Khaliullin,
         O. K. Andersen, and K. Held,
         Phys. Rev. Lett. {\bf 103}, 016401 (2009). 

\bibitem{John-arxiv} J. W. Freeland, J. Liu, M. Kareev, B. Gray,
         J.W. Kim, P. Ryan, R. Pentcheva, and J. Chakhalian,
  arXiv:1008.1518 (2010).


\bibitem{MJHan} M. J. Han, C. A. Marianetti, and A. J. Millis,
    Phys. Rev. B. {\bf 82} 134408, (2010).


\bibitem{MJHan-DMFT} M. J. Han, X. Wang, C. A. Marianetti, and A. J. Millis,
   Phys. Rev. Lett.  {\bf 107} 206804, (2011).


\bibitem{NNO-PRL} M. K. Stewart, J. Liu, M. Kareev, J. Chakhalian, and D. N. Basov,
    Phys. Rev. Lett.  {\bf 107} 176401, (2011).


\bibitem{Chaloupka-Khaliullin} J. Chaloupka and G. Khaliullin,
         Phys. Rev. Lett. {\bf 100}, 016404 (2008). 


\bibitem{LDA+U-ori}  V. I. Anisimov, J. Zaanen, and O. K. Andersen,
        Phys. Rev. B {\bf 44}, 943 (1991).

\bibitem{LDA+U-95}   A. I. Liechtenstein,  V. I. Anisimov, and J. Zaanen,
        Phys. Rev. B {\bf 52}, R5467 (1995).

\bibitem{LDA+U-rev}  V. I. Anisimov, F. Aryasetiawan, and A. I. Lichtenstein,
        J. Phys.:Condens. Matter {\bf 9}, 767 (1997);

\bibitem{Wang} X. Wang, M. J. Han, L. de' Medici, C. A. Marianetti, and A. J. Millis,
     arXiv:1110.2782 (2011).



\bibitem{troullier} N. Troullier and J. L. Martins, 
             Phys. Rev. B {\bf 43}, 1993 (1991).


\bibitem{Ozaki} T. Ozaki, Phys. Rev. B. {\bf 67}, 155108, (2003);
    T. Ozaki and H. Kino,  Phys. Rev. B {\bf 69}, 195113, (2004);
    T. Ozaki and H. Kino,  J. Chem. Phys. {\bf 121}, 10879, (2004).


\bibitem{CA} D. M. Ceperley and B. J. Alder, Phys. Rev. Lett. {\bf 45}, 566(1980);
             J. P. Perdew and A. Zunger, Phys. Rev. B {\bf 23}, 5048 (1981).


\bibitem{Han-LDA+U} M. J. Han, T. Ozaki, and J. Yu,
             Phys. Rev. B \textbf{73}, 045110 (2006).

\bibitem{Dudarev} S. L. Dudarev, G. A. Botton, S. Y. Savrasov,
                 C. J. Humphreys, and A. P. Sutton,
             Phys. Rev. B {\bf 57}, 1505 (1998).


           \bibitem{comment0} In this formalism, the effective on-site
             Coulomb interaction is represented by $U$ (on-site
             Coulomb repulsion) $-$ $J$ (Hund's interaction).


\bibitem{OpenMX}  http://openmx-square.org


\bibitem{Luders} U. L\"uders,  W. C. Sheets, A. David,  W. Prellier, and  R. Fr\'esard,
        Phys. Rev. B {\bf 80}, 241102 (2009).


\bibitem{Burton-Tsymbal} J. D. Burton and E. Y. Tsymbal,
        Phys. Rev. Lett. {\bf 107} 166601, (2011).

\bibitem{Nanda-Satpathy} B. R. K. Nanda and S. Satpathy, 
        Phys. Rev. Lett. {\bf 101} 127201, (2008).


\bibitem{Mazin} I. I. Mazin,  D. I. Khomskii,  R. Lengsdorf,
        J. A. Alonso,  W. G. Marshall, R. M. Ibberson,
        A. Podlesnyak,  M. J. Martinez-Lope,  and 
        M. M. Abd-Elmeguid,
    Phys. Rev. Lett. {\bf 98}, 176406 (2007).


\bibitem{Medarde} M. Medarde, C. Dallera, M. Grioni, B. Delley, F. Vernay,
    J. Mesot, M. Sikora, J. A. Alonso, and M. J. Martinez-Lope,
        Phys. Rev. B {\bf 80}, 245105 (2009).


   
\bibitem{RMP-actinide} K. T. Moore and  G. van der Laan,
    Rev. Mod. Phys. {\bf 81}, 235 (2009).


\bibitem{MJHan-PuAm} M. J. Han, X. Wan, and S. Y. Savrasov,
        Phys. Rev. B {\bf 78}, 060401 (2008).

\bibitem{DMFT96} A. Georges,  G. Kotliar, W. Krauth, and M. J. Rozenberg,
         Rev. Mod. Phys. {\bf 68}, 13 (1996). 

\bibitem{DMFT06} G. Kotliar, S. Y. Savrasov, K. Haule, V. S. Oudovenko, 
        O. Parcollet, and C. A. Marianetti,
         Rev. Mod. Phys. {\bf 78}, 865 (2006). 


\bibitem{DDSarma} S. R. Barman,  A. Chainani, and D. D. Sarma, 
        Phys. Rev. B {\bf 49}, 8475 (1994).


\bibitem{MJHan-slab} M. J. Han and Michel van Veenendaal,
        Phys. Rev. B. {\bf 84} 125137, (2011).



   \bibitem{comment1} We found that this feature is retained up to
     $U=9$ eV, and for the larger $U$, the charge convergency is not
     well stabilized.

   \bibitem{comment2} It should be noted that the effective U values,
     used in the self-consistent LDA$+U$ calculation and in the
     effective model DMFT one, can not be directly compared.













\end{thebibliography}
\end{document}